\documentclass[fleqn,usenatbib]{mnras}

\usepackage{newtxtext,newtxmath}
\usepackage{graphicx}	

\usepackage[T1]{fontenc}

\usepackage{graphicx}	
\usepackage{amsmath,units}	
\usepackage[normalem]{ulem}

\title[Wideband polarimetry]{Frequency- and phase-resolved polarimetry of millisecond pulsars and its application to timing}

\author[M. Cury\l{}o et al.]{
Ma\l{}gorzata Cury\l{}o,$^{1,2}$\thanks{E-mail: gosia.curylo@monash.edu}
Andrew Zic,$^{3}$
Shuangqiang Wang,$^{4,3}$
Eric Thrane,$^{1,2}$
Paul D. Lasky,$^{1,2}$
\newauthor
Jacob Cardinal Tremblay,$^{5,6}$
Zu-Cheng Chen,$^{7,8}$
Shi Dai,$^{3}$
Valentina Di Marco,$^{2,3,9}$
George Hobbs,$^{3}$
\newauthor
Agastya Kapur,$^{3,10}$
Wenhua Ling,$^{3}$
Marcus E. Lower,$^{11}$
Rami F. Mandow,$^{3,10}$
Saurav Mishra,$^{2,3,11}$
\newauthor
Daniel J. Reardon, $^{2,11}$
Christopher J. Russell,$^{12}$
Ryan M. Shannon,$^{2,11}$ and
Xing-Jiang Zhu$^{13,14}$ \\
$^{1}$ School of Physics and Astronomy, Monash University, Clayton VIC 3800, Australia\\
$^{2}$ OzGrav: The ARC Centre of Excellence for Gravitational Wave Discovery\\
$^{3}$ Australia Telescope National Facility, CSIRO, Space and Astronomy, PO Box 76, Epping, NSW 1710, Australia\\
$^{4}$ Xinjiang Astronomical Observatory, Chinese Academy of Sciences, Urumqi, Xinjiang 830011, People’s Republic of China\\
$^{5}$ Max Planck Institute for Gravitational Physics (Albert Einstein Institute), 30167 Hannover, Germany\\
$^{6}$ Leibniz Universit\"at Hannover, 30167 Hannover, Germany \\
$^{7}$ Department of Physics and Synergetic Innovation Center for Quantum Effects and Applications, Hunan Normal University, Changsha, Hunan 410081, China \\
$^{8}$ Institute of Interdisciplinary Studies, Hunan Normal University, Changsha, Hunan 410081, China \\
$^{9}$ School of Physics, University of Melbourne, Parkville, VIC 3010, Australia\\
$^{10}$ Department of Mathematics and Physical Sciences, Macquarie University, NSW 2109, Australia \\
$^{11}$ Centre for Astrophysics and Supercomputing, Swinburne University of Technology, Hawthorn, VIC, 3122, Australia \\
$^{12}$ CSIRO Scientific Computing, Australian Technology Park, Locked Bag 9013, Alexandria, NSW 1435, Australia \\
$^{13}$ Department of Physics, Faculty of Arts and Sciences, Beijing Normal University, Zhuhai 519087, China \\
$^{14}$ Institute for Frontier in Astronomy and Astrophysics, Beijing Normal University, Beijing 102206, China \\
}

\date{Accepted XXX. Received YYY; in original form ZZZ}

\pubyear{2015}

\begin{document}
\label{firstpage}
\pagerange{\pageref{firstpage}--\pageref{lastpage}}
\maketitle


\begin{abstract}
Pulsar timing is used for a variety of applications including tests of fundamental physics, probing the structure of neutron stars, and detecting nanohertz gravitational waves. Development of robust methods and generation of high-quality timing data is therefore of utmost importance. In this paper, we present a new technique for creating high-fidelity templates that can be used to measure the pulse times of arrival with significantly increased precision compared to existing methods. Our framework makes use of all available polarimetric information to generate frequency-dependent models of pulse-shape evolution of all four Stokes parameters. 
We apply this method to millisecond pulsars observed by the Parkes Pulsar Timing Array and show that it results in timing measurement uncertainties reduced up to $\sim$20-30\%. We also present, for the first time, phase- and frequency-resolved polarimetric measurements of millisecond pulsars observed with the Parkes Murriyang ultra-widebandwith-low receiver. The data, plots and the code underlying this analysis are made publicly available.  
\end{abstract}




\section{Introduction}
Millisecond pulsars are exceptional clocks with which to carry out precision timing.
The best-timed pulsars such as PSR J1909--3744 yield post-fit root-mean squared residuals $\lesssim\unit[100]{ns}$.
Such low-noise pulsars enable precision tests of fundamental physics \citep{Fonseca21, Podsiadlowski2005,Weisberg2010} and measurements of nanohertz gravitational waves \citep{Sazhin78,Detweiler79}, the first evidence for which has recently been put forward
by several collaborations \citep{Agazie2023_EvidenceGW,Reardon2023_PPTA,Antoniadis2023_EPTA_DR2,Xu2023_CPTA_DR1, Miles25}.

The key concept underpinning pulsar timing is the fact that the pulse profiles of millisecond pulsars averaged over many rotational periods are remarkably stable. 
Pulsar timers can therefore use matched filtering \citep{Taylor92} to determine the pulse times of arrival (TOAs). Obtaining accurate TOAs is of utmost importance, with more accurate templates helping reduce uncertainty.
Misspecified templates can lead to bias, for example, mimicking the signature of gravitational waves~\citep{Straten13,Lentati16}. 

The accuracy and precision of pulsar timing depends on factors both intrinsic and extrinsic to the pulsars, which make the template-matching process quite nuanced. 
First, profile shapes evolve in radio frequency $\nu$. 
Matching data with a template derived from a mean profile (averaged over $\nu$) can therefore introduce bias. 
This bias can be mitigated in two ways.
One option is to divide the band into sub-bands in order to measure sub-banded TOAs, each with its own template.
The other approach is to model out radio-frequency-dependent delays downstream in the analysis, i.e., by introducing a number of constant offsets into the timing model \citep{NG15}.

The stability of the pulse profiles is directly related to the state of pulsar's magnetosphere and emission mechanisms. In particular, processes such as jitter, nulling, mode changing and discrete state changes, all cause pulse shape variations \citep{Wang07, Oslowski13, Shannon14, Miles22, Mandow25}. Detailed polarimetry provides a means to understand these effects by probing the pulsar geometry and magnetosphere.
In this way polarimetry is therefore essential for developing and optimising pulsar timing strategies. 

One of the most significant improvements in timing methods was presented by \cite{Straten06, Straten13}, which developed the new calibration technique of \textit{measurement equation template matching} (METM) and incorporated polarisation information into the TOA generation via \textit{matrix template matching} (MTM), as opposed to the standard, scalar analysis which uses only total intensity profiles. Improvements in TOA precision come from two main reasons. Firstly, polarisation profiles are often characterised by much sharper features which provide additional information from higher harmonics. Secondly, MTM explicitly models the suboptimal receiver response to the polarised waves and thus reduces instrumental errors. Utilisation of both techniques has recently been demonstrated to be highly beneficial \citep{Rogers24, Guillemot25}.

Finally, the advent of ultra-wide-bandwidth radio observations marks another important milestone towards improving the precision of pulsar timing. There are currently two instruments allowing for such observations: the ultra-broadband receiver at the Effelsberg $\unit[100]{m}$ Radio Telescope,\footnote{https://www3.mpifr-bonn.mpg.de/staff/pfreire/BEACON.html} covering frequency range of 1.3--6.0~GHz, and the ultra-wide-bandwidth low (UWL) receiver at the 64~m Parkes Radio Telescope \textit{Murriyang} \citep{Hobbs20} spanning 0.7--4~GHz. A third such instrument is being commissioned at the 100~m Green Bank Telescope \citep{GBT23}. 
Benefits of wide instantaneous frequency coverage include: increased signal-to-noise ratio (S/N), improved measurements on frequency-dependent processes, radio-frequency-interference (RFI) mitigation, and simple, consistent instrumental setup. On the other hand, wider bands significantly enhance all shortcomings of standard timing methods arising from the heuristic approach to handling frequency-dependent processes. To address these, \cite{Liu14} and \cite{Pennucci14} introduced a new wideband timing technique that models the profile shape changes in frequency and provides a two-dimensional (phase, $\nu$) template. Many pulsar timing array collaborations have already started shifting towards these new techniques \citep{NG_WB21, Tarafdar22, Sharma22, Curylo23, Agazie2023_EvidenceGW, Zic23, Miles25}. 

Previous multi-frequency observations have been used to investigate important concepts such as emission mechanisms, emission height in the magnetosphere and beam models mainly based on the slow pulsar population (e.g., \citealp{Cordes78, Rankin83, Lyne88, Kijak03, Johnston08}). Moreover, such studies suffer from challenges due to non-contiguous frequency coverage, the usage of different receiving systems, and data obtained at different epochs. Utilization of truly broadband instruments, such as the UWL receiver, will be invaluable to fill in the gaps, reducing systematic errors from the use of disjoint data sets. The remarkable capabilities of the UWL receiver have recently been demonstrated in wideband studies of slow pulsars dedicated to beam geometries \citep{Jaroenjittichai25}, modelling of the polarised emission \citep{Oswald23b}, and population statistics \citep{Johnston21, Oswald23}. 

In this work, we build on \cite{Straten06} and \cite{Pennucci14}, developing a technique to obtain \textit{polarimetric portrait templates}, i.e., wideband, two-dimensional templates for all four Stokes parameters. By including all available information, polarimetric portrait templates have the potential to produce more refined timing measurements than previously possible. These improved measurements can then be used to provide reduced-noise measurements for a myriad of studies, including gravitational-wave astronomy. Moreover, our new template-generation process makes the high-quality polarisation information readily available. This includes phase- and frequency-resolved polarisation profiles, fractions, position angles, flux densities, and spectral indices.  

We apply our new formalism to millisecond pulsars observed as part of the Parkes Pulsar Timing Array with the UWL receiver. We present wideband phase- and frequency-resolved polarimetric properties of these objects and compare them with previous multi-frequency observations \citep{Dai15} and the population of slow pulsars \citep{Oswald23}. We employ polarimetric portrait templates to generate a new timing release for the PPTA (Wang et al., in preparation) and gravitational-wave analysis (PPTA collaboration, in preparation). We make publicly available all of our results---including average portraits, resolved polarisation profiles and analysis code as supplementary material online. 

The remainder of the paper is organised as follows.
In Section~\ref{method}, we give a pedagogical introduction to pulse portraiture and describe our methodology for the construction of polarimetric portrait templates.
In Section~\ref{data}, we describe the data used in the analysis. 
In Section~\ref{results}, we present the wideband polarimetric properties of analysed pulsars. Concluding remarks are provided in Section~\ref{conclusions}.

\section{Method}\label{method}
\subsection{Useful definitions}

Radio waves interact with Galactic free electrons and magnetic fields which introduce radio-frequency-dependent delays and Faraday rotations to the observed pulses. The quantities that describe these are the dispersion measure (DM) and the rotation measure (RM). The DM is defined as the free electron column density $n_{\rm e}$ along the line of sight $l$:
\begin{equation}\label{eq:DM}
\rm DM \equiv \int dl \, n_e .
\end{equation}
The DM describes a characteristic delay of radio waves proportional to $\nu^{-2}$, where lower-frequency radio waves arrive later than higher-frequency waves. If not properly corrected, the dispersion smears the profile, changing its sharpness, width and detectability.
Disentangling profile shape changes due to DM versus changes that are intrinsic frequency evolution remains one of the most pressing unsolved problems in pulsar timing. 

The RM characterises the Faraday rotation of the linearly polarised emission due to interaction with intervening magnetic fields.
It is given by
\begin{equation}\label{eq:2}
\rm  RM = \frac{\beta}{\lambda^2}, 
\end{equation}
where $\lambda$ is the wavelength of the radio waves and 
\begin{equation}
\beta = \frac{e^3\lambda^2}{2\pi m_e^2 c^4} \int dl \, n_e(l) B_{||}(l)  ,
\end{equation}
is the angle of the rotation with $e$ and $m_e$ the electron charge and mass, respectively, $c$ is the speed of light, $B_{||}$ is the component of the magnetic field parallel to the line of sight. 
Such rotations along with Faraday conversion and generation of orthogonal polarisation modes can be introduced not only by the Galactic magnetic fields but also the field in the pulsar's magnetosphere as well as Earth's ionosphere.

The polarisation state of the pulsar emission can be described by the Stokes vector \citep{Stokes1851, Straten10}
\begin{equation}
    \mathbf{S} =
    \begin{bmatrix}
        {\rm I} \\
        {\rm Q} \\
        {\rm U} \\
        {\rm V}
    \end{bmatrix}, 
\end{equation}
where I is the total intensity, Q and U together describe linear polarisation ${\rm L} \equiv \sqrt{{\rm U}^2 + {\rm Q}^2}$, and V is the circular polarisation. However, due to imperfections of the observing systems, the measured vector $\widehat{\mathbf{S}}$ will always be different than the true vector $\mathbf{S}$. The difference can be accounted for with the M{\"u}ller matrix $\mathbf{M}$, which characterises the receiver feed rotations, gain, and polarisation leakage, and allows for the transformation between the true and measured values as $\widehat{\mathbf{S}} = \mathbf{M~S}$ \citep{Heiles01}.

\subsection{Basics of pulse portraiture}
We build on previous work by \cite{Pennucci14,Pennucci19}, which describes a method for the creation of frequency-dependent total intensity portraits and templates.\footnote{This method is implemented in the publicly available code, \textsc{PulsePortraiture}; https://github.com/pennucci/PulsePortraiture.}
The first step is to prepare a two-dimensional portrait from a collection of one-dimensional \textit{profiles}, each of which describes the pulsar intensity as a function of phase at some fixed radio frequency. 
The average portrait is obtained by averaging ${\cal O}(10^1 - 10^2)$ epochs that have been aligned in phase and dedispersed using Eq.~\ref{eq:DM}; that is, the portrait is corrected to undo the effect of DM.
Next, the averaged portrait is decomposed into principal components $\hat{e}_i$ with principal component analysis. 
The components with the highest S/N represent those which are most important in representing the frequency evolution of the pulse shape. 
Of the $n_\text{chan}$ components (where $n_\text{chan}$ is the number of frequency channels), typically only $n_\text{eig}$ = a few are needed to provide an adequate representation of the data.
We call these $n_\text{eig}$ components used to reconstruct the data \textit{eigenprofiles}. 

Each profile from the average portrait is then subtracted from the mean profile and projected onto the eigenprofiles. These projections result in $n_{\rm eig}$ coordinate functions that encode the frequency evolution of the pulse shape.  
The final model is obtained by fitting a spline function to each of the coordinate functions allowing for a smooth, continuous interpolation across frequencies. The model allows for generation of a noise-free template that we write as
\begin{equation}
T(\nu) = \sum^{n_{\rm eig}}_{i=1} S_i(\nu) \, \hat{e}_i + \tilde{p} .
\end{equation}
Here, $\hat{e}_i$ are eigenprofiles, $S_i(\nu)$ are the spline functions, and $\tilde{p}$ is the mean profile.

In this work, we build on this formalism to provide polarimetric portrait templates with the following outputs:
\begin{itemize}
\item generation of frequency-resolved portraits of all four Stokes parameters,
\item calculation of linear polarisation from Stokes Q and U,
\item calculation of polarisation fractions and position angles resolved in phase and frequency as well as phase-averaged,
\item principal component and spline models for each Stokes parameter, and
\item a final polarimetric portrait template based on the polarimetric spline models.
\end{itemize}

Linear polarisation is calculated as in \cite{Muller17}
\begin{equation}
{\rm L} = \sqrt{{\rm Q}^2 + {\rm U}^2} = \sqrt{({\rm Q}_T+{\rm Q}_N)^2 + ({\rm U}_T + {\rm U}_N)^2}, 
\end{equation}
where ${\rm Q}_T$, ${\rm U}_T$ correspond to astrophysical polarised emission, whereas ${\rm Q}_N$ and ${\rm U}_N$ are the receiver noise contributions. The presence of noise introduces a positive bias into the measured values of L, which we remove by applying corrections described by \cite{Muller17}. We correct an analogous bias in |V| following \cite{Oswald23}:
\begin{equation}
|V|^{\rm corr} =
\begin{cases}
|V| - \sigma_{\rm I} \sqrt{\tfrac{2}{\pi}}, & \text{if } |V| > \sigma_{\rm I} \sqrt{\tfrac{2}{\pi}} \\[6pt]
0 & \text{otherwise}
\end{cases} ,
\end{equation}
where $\sigma_{\rm I}$ is off-pulse noise measured in Stokes $I$. 

Position angle (PA) is defined as:
\begin{equation}
\rm \psi = 0.5~\rm~tan^{-1}(U/Q),
\end{equation}
with uncertainty in the $i^\text{th}$ frequency channel $\sigma_{\rm \psi_i} = \sigma_{\rm I,i}/2L_{\rm i}$. However, due to the above-mentioned noise bias, PA uncertainty in the low signal-to-noise regime (i.e., where $L_{\rm i}/\sigma_{I,i}<10$) has to be calculated numerically as described in \cite{Naghizadeh93} and \cite{Everett01}.

We calculate linear, net circular, and absolute circular polarisation fractions, denoted $l$, $v$, and $|v|$, respectively. For each, we calculate them resolved in phase and frequency, as well as phase-averaged. In the case of the latter, we sum contributions only from on-pulse bins which were determined with the minimum baseline estimator available in \textit{psrstat} package.\footnote{https://psrchive.sourceforge.net/manuals/psrstat/algorithms/}
Similarly to \cite{Oswald23}, we implement two types of phase-averaged fractions as defined in Eq.~\ref{eq:fractions}.
In one version, we first calculate on-pulse means of the Stokes parameters and then polarisation fractions (marked with an asterisk). In the second version, the noise-bias corrected L and V are calculated first and then averaged in phase to form fractions (marked with a bar). 

\begin{equation}\label{eq:fractions}
  \begin{aligned}
    l^*            &= \frac{\sqrt{(\Sigma {\rm Q})^2 + (\Sigma {\rm U})^2}}{\Sigma {\rm I}}, \\[6pt]
    |v|^*          &= \left| \frac{\Sigma {\rm V}}{\Sigma {\rm I}} \right|, \\[6pt]
    \overline{l}   &= \frac{\Sigma\!\left( L^{\rm corr} \right)}{\Sigma {\rm I}}, \\[6pt]
    \overline{|v|} &= \frac{\Sigma\!\left( |V|^{\rm corr} \right)}{\Sigma {\rm I}}. \\[6pt]
  \end{aligned}
\end{equation}

Each of these sets of measurements carry different information and so both are included in our code. However, as barred fractions are the ones usually being reported in the literature and are less affected by averaging effects (note that PA and $V$ may change sign across phase), all of the plots and tables presented here contain only those. 

We calculate uncertainties on polarisation fractions as standard errors $\sigma$ of the on-pulse means $\sigma/\sqrt{N_{\rm on}}$, where $N_{\rm on}$ is the number of on-pulse bins. 
Finally, principal component and spline modelling are performed separately for each of the Stokes parameters following the standard wideband technique described above.  

\subsection{Flux and spectral indices}
In order to minimise the effect of scintillation, we first measure flux in each individual observation averaged in phase over the entire Stokes $I$ profile and then calculate the mean with uncertainty given by:
\begin{equation}
S^{\rm RMS}/(N_{\rm obs}-1)^{1/2},
\end{equation}
where $S^{\rm RMS}$ and $N_{\rm obs}$ are the square root of the variance and the number of observations, respectively. 
We also calculate phase-resolved spectral indices, which are also measured as mean of Stokes $I$ across all individual observations. Our flux measurements are obtained with all available observations for which we were able to get reliable DM and RM measurements. 

\section{Data}\label{data}
The data set consists of observations obtained with the UWL receiver between November 2018 and January 2025 (MJDs 58425 -- 60694) as part of the Parkes Pulsar Timing Array project. Data are taken with a typical cadence of three weeks and nominal integration time of 3840~s per pulsar. Calibration of flux and polarisation is performed with three sources: a) a noise diode (observed right before and after each observation), b) primary flux calibrator PKS B0407-658 or PKS B19324-638, used for rescaling flux to astrophysical units (observed once every observing session), and c) PSR J0437--4715 to allow for receiver ellipticity corrections via METM (observed rise-to-set every few weeks). The raw data are coherently dedispersed and folded into archives with 3328 frequency channels and 1024 phase bins. Each observation then undergoes automated two-stage RFI excision and calibration with standard \textit{pac} routines available with \textsc{PSRCHIVE} (a detailed description of UWL data processing can be found in \citealp{Zic23}).

We pre-process the data by correcting them for RM and DM. In order to correct for DM, we generate frequency-resolved timing residuals with \textsc{TEMPO2} using templates from the PPTA third data release \citep{Curylo23, Zic23}. We fit for DM per observation, fixing all other timing parameters. 
The RM in each observation is fit and corrected with the \textit{rmfit} routine from \textsc{PSRCHIVE}. We use the brute-force method of searching for best RM, which maximises the amount of linear polarisation, and fit the RM spectrum with a Gaussian.
The final set of observations is obtained by filtering data corrupted due to excessive RFI, instrumental errors and failed DM and RM fits.

\section{Results and discussion}\label{results}
We first discuss our templates and general properties of pulsars in terms of their polarisation profiles, fractions and fluxes. Later, in Appendix~\ref{appendix:A}, we provide a short summary of each pulsar, highlighting interesting features. Plots presented below show only a subset of our results. The full data set is available online as supplementary material. 

\subsection{Polarised portraits and templates}
Our first main result is a set of high-quality portraits (averaged and aligned observations) and templates (derived from the spline models) for each pulsar. These are available publicly together with the code so that the results presented here can be
\begin{enumerate}
\item easily reproduced,
\item modified or extended,
\item recreated for other pulsars, and 
\item used for further analyses.
\end{enumerate}
Each portrait has 832 frequency channels and 1024 phase bins. 

We test the performance of the new templates running an initial timing campaign of the PPTA pulsars. We compare the results obtained with the scalar template matching (STM), which uses only total intensity profiles, to results obtained with MTM utilising polarimetric information. Mean uncertainties of the TOAs and root-mean-square of the timing residuals of the three pulsars which improved the most are presented in Tab~\ref{tab:mtm}. Improvement for PSR J0437--4715 (the brightest pulsar in the sample) mostly comes from correcting for polarisation calibration errors, whereas the other two pulsars are highly polarised and so polarimetric timing is supported by much more information than STM. Less improvement is expected for fainter pulsars with lower polarisation fractions. The full timing data set of all PPTA pulsars will be presented in Wang et al. (in preparation). 

\begin{table}
\centering
\begin{tabular}{|l|c|c|c|c|}
\hline
\textbf{Pulsar} & \multicolumn{2}{c|}{\textbf{STM}} & \multicolumn{2}{c|}{\textbf{MTM}} \\ \hline
 & TOA & RMS & TOA & RMS \\ \hline
J0437--4715 & 0.104 & 0.147 & 0.068 & 0.097 \\ \hline
J1022+1001 & 2.881 & 1.991 & 2.297 & 1.499 \\ \hline
J1024--0719 & 2.657 & 1.597 & 2.013 & 1.146 \\ \hline
\end{tabular}
\caption{Comparison of timing results with standard total intensity template-matching (STM) and MTM. Columns with TOA and RMS correspond to mean TOA uncertainty and root-mean-square of the residuals, respectively.}
\label{tab:mtm}
\end{table}

\subsection{Polarisation profiles, fractions and position angles}
\begin{figure*}
\centering
\includegraphics[scale=0.9]{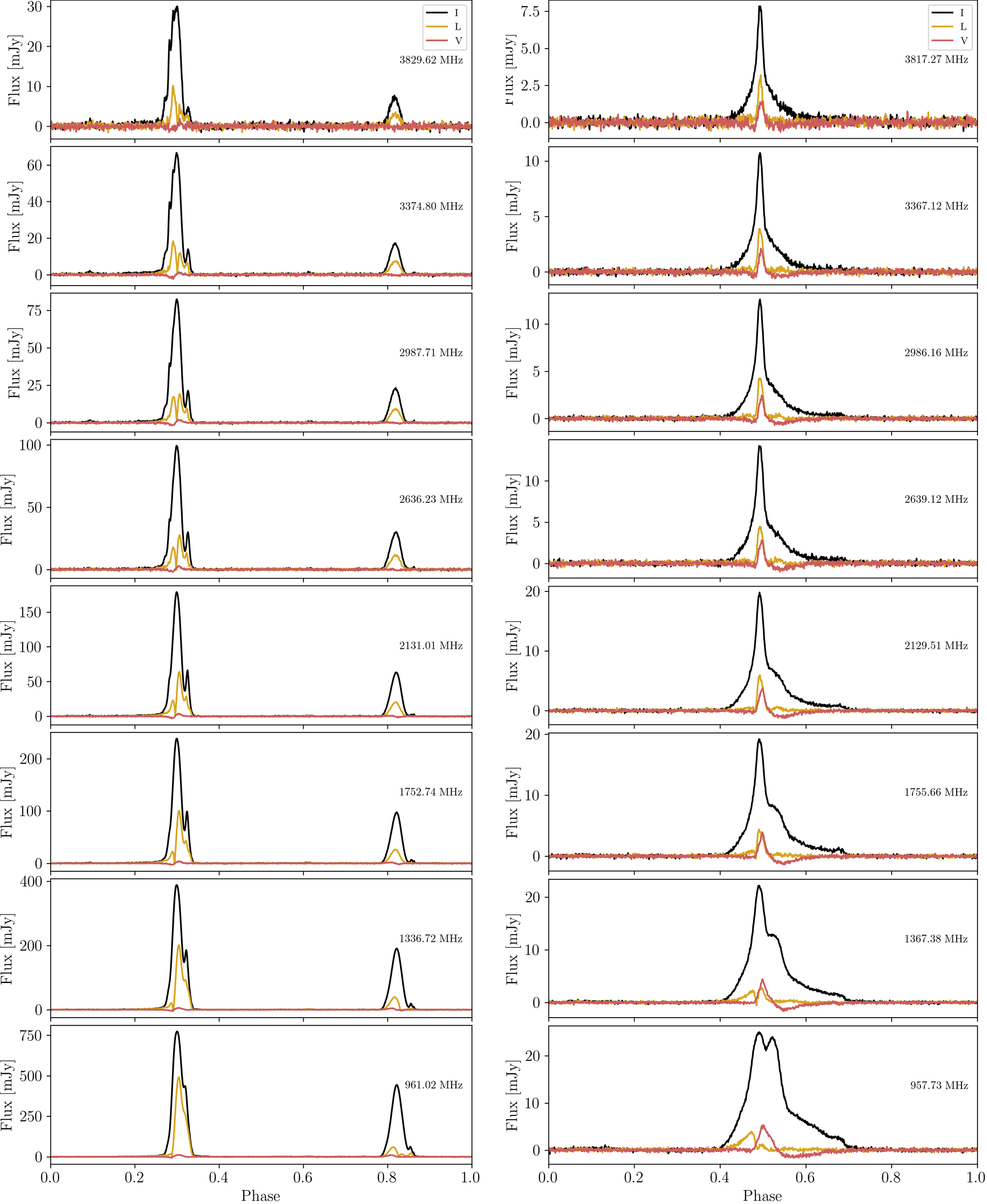}
\caption{Polarisation profiles of I (black), L (gold) and V (red) in 8 sub-bands. Left is PSR  J1939+2134;  right is PSR J2051--0827.}
\label{fig:8prof}
\end{figure*}

\begin{figure*}
\centering
\includegraphics[scale=1]{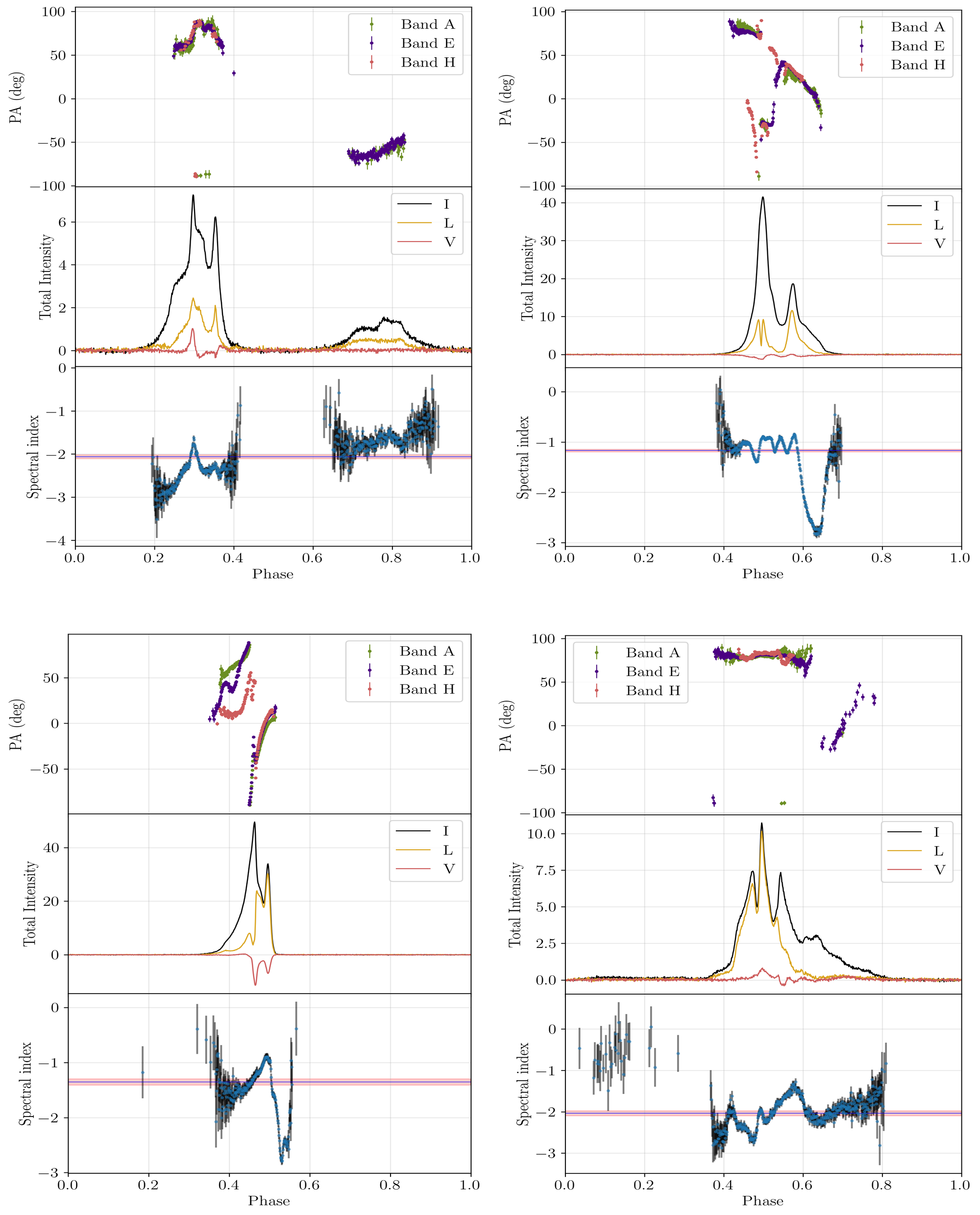}
\caption{Fully band-averaged polarisation profiles, PAs and phase-resolved spectral indices for four pulsars (left to right and top to bottom): PSR J0030+0451, PSR J0125--2327, PSR J1022+1001, PSR J1024--0719. The purple shaded line in the bottom plots shows the phase averaged value of the spectral index.}
\label{fig:band0}
\end{figure*}

\begin{figure*}
\includegraphics[scale=1.1]{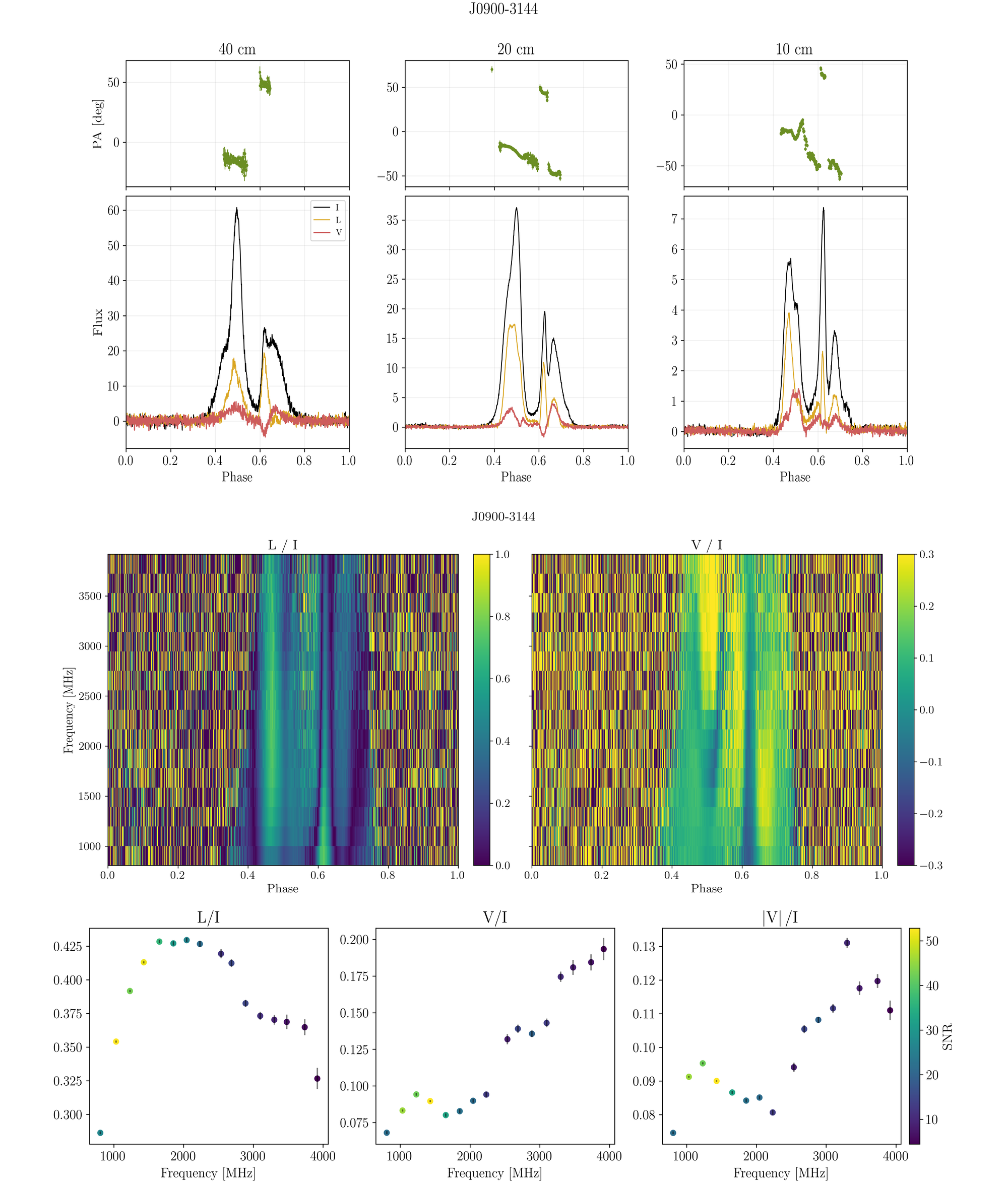}
\caption{Results for PSR J0900--3144. Top: polarisation profiles and PAs in three bands, middle: waterfall plots of phase- and frequency-resolved polarisation fractions (16 frequency channels), bottom: phase-averaged polarisation fractions (16 frequency channels).}
\label{fig:J0900}
\end{figure*}

Pulsar profiles can be characterised by a number of \textit{profile components}---distinct peaks in the pulse profile.\footnote{In order to avoid confusion, we distinguish between ``profile components'' (features in the profile shape) and ``principal components''(used the construction of the eigenprofiles).}
In order to find balance between maintaining high S/N and time resolving the frequency evolution of profile components, we present polarisation profiles from portraits sub-banded into eight frequency channels as shown in Fig.~\ref{fig:8prof}.
For the purpose of comparison with previous works, we also show results averaged over the entire UWL band (Fig.~\ref{fig:band0}) and at three frequency bands: A (40 cm, bandwidth of 64~MHz), E (20 cm, bandwidth of 256~MHz), and H (10 cm, bandwidth of 1440~MHz) (top panel of  Fig.~\ref{fig:J0900}). All profiles are centred at phase~=~0.5, unless they are wrapped at the edges. In these cases, we rotate the profile in the plots (portrait remains centred at 0.5). 

We calculate polarisation fractions in 8, 16 and 32 sub-bands, both resolved and averaged in phase. An example ``waterfall plot'' of phase-frequency resolved, and phase-averaged $L/I$ and $V/I$ are shown in the middle and bottom panels of Fig.~\ref{fig:J0900}. In contrast to \cite{Oswald23}, we do not remove the off-pulse bins in the waterfall plots for better interpretability. While data with 8 sub-bands is characterised by higher S/N and lower uncertainties, higher resolution with 16 or 32 sub-bands better illustrates the smooth changes in polarisation.

The majority of pulsars in our sample have complex, multicomponent pulse profiles, often dramatically evolving with frequency. Various profile components have different polarisation fractions, and their shapes can evolve differently with frequency. 
In some pulsars, one of the profile components stays highly polarised across the band while the rest depolarise, e.g., PSR J0614--3329, PSR J0711--6830, and PSR J1022+1001.

Eighteen pulsars have disjoint profile components or interpulses where the different profile components do not overlap.\footnote{These include PSR J0030+0451, PSR J0614--3329, PSR J1045--4509, PSR J1125--6014, PSR J1545--4550, PSR J1603--7202, PSR J1730--2304, PSR J1744--1134, PSR J1824--2352A, PSR J1832--0836, PSR J1857+0943, PSR J1902--5105, PSR J1909--3744, PSR J1939+2134, PSR J2051--0827, PSR J2145--0750, PSR J2241--5236.}
Often, disjoint profile components have a high degree of linear polarisation (especially at the lowest frequencies), sometimes maintained across the whole band where they are visible.\footnote{See, e.g., PSR J1545--4550, PSR J1603--7202, PSR J1730--2304, PSR J1824--2452A, and PSR J1832--0836.} This is particularly interesting in light of the recent results from \cite{Kramer25}, who argue that radio emission can in fact be produced beyond the light cylinder along with gamma ray emission. Properties of these light-cylinder profile components are expected to be different from the ones produced in a classical way above the polar cap. A detailed study of this sample of PPTA pulsars will be presented in the follow-up paper.

The majority of PPTA pulsars have PAs with many orthogonal jumps and generally high level of variability that cannot be explained well with the rotating vector model (RVM; \citealp{RVM}). For many pulsars, the the shape of PA curves evolve with radio frequency and they are also generally flatter at phases corresponding to higher L fractions with less radio-frequency evolution (see, e.g., PSR J1024--0719 in the bottom right panel of Fig.\ref{fig:band0}). What is striking is that PA curves for all pulsars are almost perfectly aligned in phase (see Fig.\ref{fig:band0}). It is commonly believed that the shift in PAs is related to the height of emission \citep{Blaskiewicz91}. Our data therefore shows that radio emission of millisecond pulsars might be generated at an extremely narrow, nearly identical height range. Further, it would mean that radius-to-frequency mapping \citep{Cordes78} which is applicable for the population of slow pulsars, does not hold in the case of millisecond pulsars. 

For most pulsars, V changes sign across phase, and also radio frequency. All pulsars have detectable $V$, although usually low (except from PSR J0711--6830, PSR J1017--7156, PSR J1045--4509, PSR J1603--7202, PSR J1730--2304, PSR J2129--5721). One of the pulsars, PSR J1902--5105, is also interesting with almost no L, but prominent V. 

Most pulsars depolarise with increasing radio frequency, but there are a few exceptions. Some fractions of L and V are unimodal functions of frequency.\footnote{See, e.g., PSR J0125--2327, PSR J0437--4715, PSR J0614--3329, PSR J0900--3144, PSR J1017--7156, PSR J1125--6014, PSR J1545--4550, PSR J1600--3053, PSR J1603--7202, J1643--1224, J1730--2304, PSR J2241--5236.}
However, minima/maxima may occur at different frequencies in L and V and directions also may be different. An example of such unimodal frequency evolution of phase-averaged polarisation fraction is shown in the bottom panel of Fig.~\ref{fig:J0900}. 
Some pulsars show opposite trends in different profile components; see, e.g., PSR J1939+2134 in the left panel of Fig.~\ref{fig:8prof}, where linear polarisation fraction is increasing with frequency in the profile component on the left and decreasing in the component on the right. 

We report previously detected micro-components in profiles of PSR J0900--3144, PSR J1024--0719, PSR J1545--4550, PSR J1603--7202, PSR J1909--3744, and PSR J1939+2134. We note that there is a very faint and previously unreported component in PSR J1909--3744 following the main component. 
However, inspecting multi-frequency profiles from \cite{Dai15}, it can be seen that the component was visible at the time, though missed, likely because of a very bright neighbouring main component. There is a very faint micro-component also in PSR J1600--3053 at phases $\sim$0.15-0.25 visible only in the lowest frequency bin. 

Generally, profile-shape complexity decreases with frequency, but there are exceptions where it increases or does not change.\footnote{See, e.g., PSR J0900--3144, PSR J1125--6014, PSR J1600--3053, PSR J1603--7202, PSR J1643--1224, PSR J1713+0747, PSR J1730--2304, PSR J1939+2134, PSR J2124--3358, PSR J2145--0750.}
Pulsars with a single, clear profile component in total intensity can have complex shapes in L and V, which evolve significantly with frequency.\footnote{See, e.g., PSR J1017--7156, PSR J1643--1224, PSR J1857+0943, PSR J1705--1903, PSR J1909--3744, PSR J1939+2134, PSR J2241--5236).}

Phase- and frequency-resolved polarisation fractions reveal presence of characteristic arcs in many pulsars (as an example see PSR J0900--3144 in Fig.\ref{fig:J0900}). These arcs reflect smooth and astoundingly symmetrical coevolution in frequency of polarisation fractions and profile shapes. However, arcs in fractions of L and V seem to be independent of each other. 

Some of our profiles (e.g., PSR J0613--0200, PSR J1017--7156, PSR J1939+2134) are sharper than previously published results \citep{Dai15}, and reveal new features at the trailing edges of the main profile components (see, e.g., PSR J1939+2134 in the left panel of Fig.\ref{fig:8prof}). This is due to careful corrections of DM we apply in preprocessing of the data which reduces the profile smearing. 

\begin{figure*}
\centering
\includegraphics[scale=0.52]{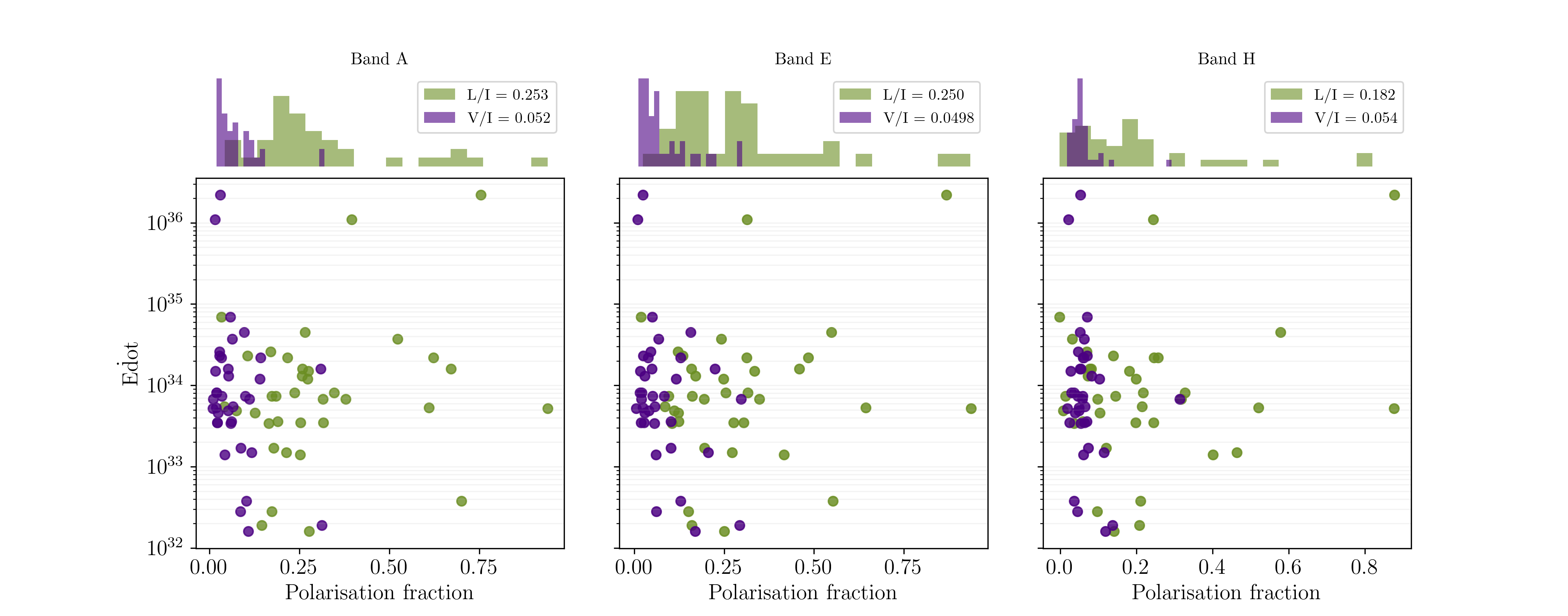}
\caption{Linear and absolute circular polarisation fractions as a function of spin-down energy $\dot{E}$ in three bands. The legend shows median values across all pulsars.}
\label{fig:fra_evol}
\end{figure*}

Wideband study of slow pulsars showed that pulsars with higher spin-down energy $\dot{E}$ are characterised by simpler profiles, less profile evolution, and higher fractions of L~\citep{Oswald23}. This is supported by results from~\cite{Kara07} who argue that these properties are related to a smaller range of emission heights, and also because  magnetospheres of millisecond pulsars are much smaller due to shorter rotational periods \citep{Kramer98}. This could result in generally more complex and evolving profile shapes. In our sample of pulsars, $\dot{E}$ covers a relatively broad range of $\log_{10}[\dot{E}/(\rm erg\,s^{-1})]~=~32~-~34$. We see that pulsars with the largest spin-down energies ($\dot{E}>2\times10^{34}\,\rm erg\,s^{-1}$) indeed are the ones with the simplest profiles (except PSR J1824--2452A). Pulsars with the lowest $\dot{E}$ all have complex multi-component profiles. We do not see a correlation between $\dot{E}$ and polarisation fractions (Fig.~\ref{fig:fra_evol}), however this might be due to our small sample size. Moreover, median polarisation fractions across all pulsars are rather constant in the three bands (there is a slight decrease of linear polarisation and increase of absolute circular polarisation fraction). On average, PPTA pulsars are linearly polarised in $\sim$20\% and circularly polarised in $\sim5$\%, which is consistent with the population of slow pulsars \citep{Oswald23,Posselt}.

\subsection{Flux and spectral indices}
\begin{figure*}
\centering
\includegraphics[scale=0.9]{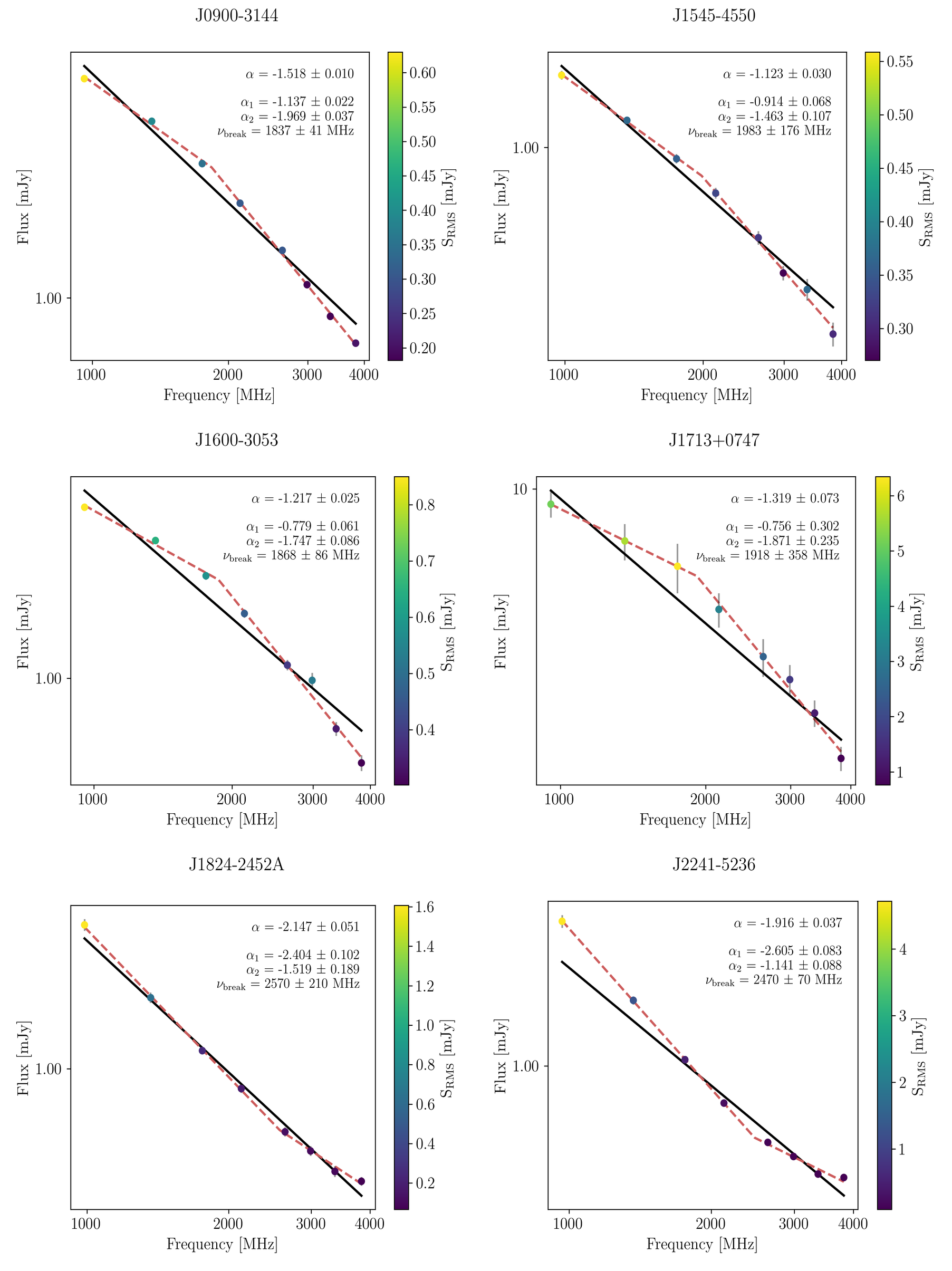}
\caption{Flux density spectra for PSR J0900--3144, PSR J1545--4550, PSR J1600--3053, PSR J1713+0747, PSR J1824--5236 and PSR J2241--5236. Colour scale represent the level of flux variations across all observations.}
\label{fig:flux2}
\end{figure*}

A majority of our pulsars have spectra which are well described by power laws and in general are consistent with previous measurements from \cite{Dai15, Gitika23}. We confirm high frequency turn-over in PSR~J2241--5236, and see a similar flattening of the spectral index at high frequencies for PSR J0030+0451, PSR J1017--7156, and PSR J1944+0907. However, usually the flattening occurs just in the last frequency bin, so we do not find it conclusive. There are six pulsars where spectra are best described by a broken power law (PSR J0900--3144, PSR J1545--4550, PSR J1600--3053, PSR J1713+0747, PSR J1824--2452A and PSR J2241--5236). These pulsars are shown in Fig.~\ref{fig:flux2}. When we compare our results to the results from \cite{Dai15, Gitika23}, which are best fitted with single power laws, we see that they are consistent, however, larger frequency coverage and resolution of the UWL allows for much more precise characterisation of the spectra. 

Phase-resolved spectral indices show significant variations of two (often mixing) types: a) fast oscillations and b) sharp drop downs or rises (a good example is seen in PSR J0125--2327 in Fig.~\ref{fig:band0}). Oscillations generally correspond to more complicated profile shapes, whereas smooth changes are seen in simpler profiles (eg. PSR J1045--4509). Often, sharp drop downs or rises are associated with parts of the profile that most significantly evolve with frequency. 

We list all pulsars in Tab.~\ref{tab:tab_all} along with their rotational periods P, median DM and RM values obtained from fitting, and flux densities and spectral indices.

\begin{table*}
\centering
\caption{Periods, DMs, RMs, flux densities at 1.4 GHz, and spectral indices of all analysed pulsars. The last two columns show S/N of the average portrait and the number of observations used to calculate flux. The DMs and RMs are means across the whole data span from the fitted time series. Pulsars that were absent in \protect\cite{Dai15} are denoted with an asterisk.}
\label{tab:tab_all}
\begin{tabular}{lccc|c|c|c|c|cclc}
\hline
Pulsar & P & DM & RM & $S_{1400}$ & $\alpha$ & $\alpha_2$ & $\nu_{\text{break}}$ & S/N & Nobs \\ 
       & (ms) & (pc cm$^{-3}$) & (rad m$^{-2}$) & (mJy) &  &  & (MHz) &  & \\
\hline
J0030+0451* & 4.87 & 4.33255(5) & -0.78(2) & 1.56(4) & -2.06(4) & & & 513 & 60 \\ 
J0125--2327* & 3.67 & 9.59612(1) & 3.411(7) & 3.28(7) & -1.17(2) & & & 3879 & 203 \\ 
J0348+0432* & 39.1 & 40.4640(17) & 8(1)  & 0.69(2) & -1.86(7) & &  & 367 & 59 \\
J0437--4715 & 5.76 & 9.59612(1) & 3.411(7) & 146.58(77) & -1.898(7) & &  & 20837 & 341 \\ 
J0613--0200 & 3.06 & 38.77470(2) & 17.46(1) & 2.39(3) & -1.98(3) & &  & 1352 & 133 \\ 
J0614--3329* & 3.15 & 37.05505(5) & 37.46(3) & 0.83(2) & -1.83(4) & &  & 531 & 103 \\ 
J0711--6830 & 5.49 & 18.40957(2) & 22.15(2) & 3.36(12) & -2.10(5) & &  & 2185 & 201 \\ 
J0900--3144* & 11.11 & 75.68771(3) & 83.535(5) & 3.95(2) & -1.14(2) & -1.97(4) & 1837(41) & 2006 & 131 \\ 
J1017--7156 & 2.34 & 94.214641(7) & -66.24(1) & 0.95(1) & -1.92(2) & & & 1353 & 242 \\ 
J1022+1001 & 16.45 & 10.25576(2) & 0.719(5) & 3.47(16) & -1.35(6) & & & 4385 & 81 \\ 
J1024--0719 & 5.16 & 6.48561(3) & -4.678(8) & 2.40(11) & -2.04(6) & & & 1398 & 58 \\ 
J1045--4509 & 7.47 & 58.12042(4) & 92.70(1) & 2.74(4) & -2.16(3) & & & 1927 & 94 \\ 
J1125--6014* & 2.63 & 52.93017(2) & 7.13(2) & 1.39(2) & -1.29(2) &  &  & 758 & 169 \\ 
J1446--4701 & 2.19 & 55.82405(8) & -10.8(1) & 0.58(1) & -1.74(5) & & & 414 & 75 \\ 
J1545--4550 & 3.58 & 68.38877(2) & 2.39(1) & 1.15(2) & -0.91(7) & -1.46(11) & 1983(176) & 834 & 175 \\ 
J1600--3053 & 3.60 & 52.32788(1) & -13.36(1) & 2.51(4)  & -0.78(6) & -1.75(9) & 1863(86) & 2666 & 101 \\ 
J1603--7202 & 14.84 & 38.04927(2) & 28.51(1) & 3.23(7) & -2.70(4) & & & 4301 & 173 \\ 
J1643--1224 & 4.62 & 62.39813(2) & -305.91(1) & 4.56(3) & 1.83(1) & & & 3045 & 84 \\
J1705--1903* & 2.48 & 57.50600(2) & -16.89(4) & 0.73(2) & -1.45(5) & & & 911 & 41 \\
J1713+0747 & 4.57 & 15.99096(3) & 10.173(8) & 6.69(61) & -0.76(30) & -1.87(24) & 1918(358) & 5756 & 41 \\ 
J1730--2304 & 8.12 & 9.62464(3) & -6.829(9) & 4.76(13) & -1.75(6) & & & 3017 & 65 \\ 
J1741+1351 & 3.75 & 24.19670(8) & 59.5(1) & 0.81(7) & -1.86(13) & &  & 210 & 19 \\
J1744--1134 & 4.07 & 3.13791(1) & 0.654(3) & 3.01(12) & -1.78(5) & & & 4756 & 129 \\
J1824-2452A & 3.05 & 119.90562(6) & 80.84(2) & 2.15(6) & -2.40(10) & -1.52(19) & 2570(210) & 421 & 25 \\ 
J1832--0836 & 2.72 & 28.19112(8) & 41.1(1) & 1.54(7) & -1.95(9) & & & 313 & 33 \\ 
J1857+0943 & 5.36 & 13.29868(4) & 21.64(2) & 5.34(15) & -1.66(7) & & & 1417 & 54 \\ 
J1902--5105 & 1.74 & 36.25336(6) &  24.2(3) & 1.08(3) & -2.74(7) & & & 241 & 24 \\ 
J1909--3744 & 2.95 & 10.390714(3) & -1.699(4) & 2.34(6) & -1.22(4) & & & 9131 & 232 \\ 
J1933--6211* & 3.54 & 11.51397(4) & 7.92(5) & 1.48(8) & -1.70(8) & & & 765 & 55 \\ 
J1939+2134 & 1.56 & 71.01327(4) & 6.514(6) & 12.69() & -2.69(5) & & & 2415 & 34 \\ 
J1944+0907* & 5.19 & 24.3513(2) & -38.05(8) & 3.53(30) & -3.13(16) & & & 386 & 17 \\ 
J2051--0827* & 4.51 & 20.73044(6) & -31.54(9) & 1.81(8) & -1.42(7) & & & 1155 & 44 \\ 
J2124--3358 & 4.93 & 4.59525(4) & -0.37(1) & 5.38(16) & -2.54(4) & & & 377 & 58 \\ 
J2129--5721 & 3.73 & 31.84863(3) & 20.35(1) & 1.31(4) & -2.57(5) & & & 1063 & 151 \\ 
J2145--0750 & 16.05 & 9.00016(2) & -2.368(8) & 7.91(33) & -1.93(5) & & & 5210 & 92\\ 
J2150--0326 & 3.5 & 20.6739(4) & 6.67(66) & 1.024(6) & -1.36(10) & &  & 124 & 10\\
J2241--5236 & 2.19 & 11.410450(5) & 12.44(2) & 1.95(5) & -2.61(8) & -1.14(9) & 2470(70) & 3520 & 184 \\
\hline
\end{tabular}
\end{table*}

\begin{table*}
\label{tab:tab2}
\centering
\caption{Linear, net circular and absolute circular polarisation fractions averaged over the entire UWL band (first three columns) and measured in three sub-bands at 40~cm, 20~cm and 10~cm. }
\begin{tabular}{l|ccc|ccc|ccc|ccc}
\hline
& \multicolumn{3}{c|}{Band averaged fractions} 
& \multicolumn{3}{c|}{L/I} 
& \multicolumn{3}{c|}{V/I} 
& \multicolumn{3}{c}{|V|/I} \\
& L/I & V/I & |V|/I 
& 40\,cm & 20\,cm & 10\,cm 
& 40\,cm & 20\,cm & 10\,cm 
& 40\,cm & 20\,cm & 10\,cm \\   
\hline
J0030+0451  & 0.275 & 0.009 & 0.027 & 0.253 & 0.276 & 0.198 & 0.019 & 0.003 & 0.085 &0.022 & 0.027 & 0.064 \\
J0125--2327  & 0.275 & -0.029 & 0.021 & 0.275 & 0.334 & 0.182 &-0.001 & -0.028 & -0.032 & 0.017 & 0.016 & 0.028 \\
J0348+0432  & 0.257 & -0.191 & 0.146 & 0.277 & 0.250 & 0.141 & -0.133 & -0.208 & -0.217 & 0.108 & 0.170 & 0.118 \\
J0437--4715 & 0.242 & -0.059 & 0.117 & 0.273 & 0.249 & 0.199 & -0.031 & -0.055 & -0.078 & 0.141  & 0.116 & 0.104 \\
J0613--0200  & 0.171 & 0.031 & 0.046 & 0.258 & 0.170 & 0.073 &-0.033 & 0.027 & 0.143 & 0.053 & 0.029 & 0.083 \\
J0614--3329  & 0.299 & 0.002 & 0.035 & 0.217 & 0.313 & 0.257 & 0.017 & -0.006 & 0.042 & 0.033 & 0.038 & 0.060 \\
J0711--6830  & 0.127 & -0.150 & 0.092 & 0.189 & 0.124 & 0.059 &-0.106 & -0.155 & -0.149 & 0.061 & 0.101 & 0.070 \\
J0900--3144  & 0.403 & 0.089 & 0.057 & 0.253 & 0.417 & 0.401 & 0.067 & 0.089 & 0.127 & 0.043 & 0.060 & 0.062 \\
J1017--7156  & 0.335 & -0.279 & 0.251 & 0.316 & 0.348 & 0.318 & 0.08 & -0.337 & -0.464 & 0.111 &  0.297 & 0.313 \\
J1022+1001  & 0.459 & -0.103 & 0.097 & 0.701 & 0.553 & 0.211 & -0.114 & -0.141 & -0.019 & 0.103 & 0.128 & 0.036 \\
J1024--0719  & 0.656 & 0.023 & 0.025 & 0.610 & 0.644 & 0.521 & 0.023 & 0.022 & 0.050 & 0.018 & 0.024 & 0.049 \\
J1045--4509  & 0.186 & 0.125 & 0.095 & 0.179 & 0.195 & 0.120 &0.097 & 0.122 & 0.163 & 0.088 & 0.101 & 0.073 \\
J1125--6014  & 0.260 & 0.010 & 0.0203 & 0.346 & 0.255 & 0.218 & 0.020 & 0.013 & 0.011 & 0.020 & 0.015 & 0.030 \\
J1446--4701  & 0.261 & -0.118 & 0.069 & 0.523 & 0.242 & 0.031 & -0.118 & -0.122 & -0.079 & 0.063 & 0.068 & 0.063 \\
J1545--4550  & 0.567 & -0.152 & 0.111 & 0.266 & 0.549 & 0.578 & -0.207 & -0.205 & -0.072 & 0.097 & 0.156 & 0.052 \\
J1600--3053  & 0.318 & 0.0004 & 0.025 & 0.237 & 0.316 & 0.328 & 0.010 & 0.013 & -0.025 & 0.020 & 0.020 & 0.036 \\
J1603--7202  & 0.170 & 0.270 & 0.283 & 0.145 & 0.159 & 0.208 & 0.324 & 0.267 & 0.211 & 0.313 & 0.293 & 0.138 \\
J1643--1224  & 0.164 & -0.013 & 0.073 & 0.173 & 0.161 & 0.145 & 0.065 & -0.019 & -0.051 & 0.100 & 0.083 & 0.047 \\
J1705--1903  & 0.203 & 0.014 & 0.022 & 0.183 & 0.252 & 0.080 & 0.014 & 0.012 & 0.022 & 0.039 & 0.014 & 0.038 \\
J1713+0747  & 0.285 & -0.014 & 0.020 & 0.317 & 0.305 & 0.245 & -0.018 & -0.016 & -0.003 & 0.022 & 0.019 & 0.024 \\
J1730--2304  & 0.321 & -0.203 & 0.178 & 0.214 & 0.272 & 0.464 & -0.150 & -0.231 & -0.126 & 0.117 & 0.206 & 0.115 \\
J1741+1351  & 0.132 & -0.010 & 0.031 & 0.106 & 0.135 & 0.139 & 0.037 & 0.009 & -0.064 & 0.028 & 0.025 & 0.071 \\
J1744--1134  & 0.927 & 0.003 & 0.008 & 0.939 & 0.937 & 0.876 & 0.015 & 0.004 & 0.004 & 0.009 & 0.005 & 0.019 \\
J1824--2452A & 0.844 & -0.001 & 0.025 & 0.754 & 0.869 & 0.877 & 0.003 & -0.003 & 0.019 & 0.030 & 0.023 & 0.054 \\
J1832--0836  & 0.152 & -0.031 & 0.047 & -- & 0.159 & 0.077 & -- & -0.027 & -0.041 & -- & 0.049 & 0.052 \\
J1857+0943  & 0.118 & 0.005 & 0.031 & 0.126 & 0.122 & 0.104 & 0.018 & 0.002 & 0.013 & 0.024 & 0.029 & 0.040 \\
J1902--5105  & 0.024 & 0.066 & 0.054 & 0.033 & 0.018 & -0.001 & 0.043 & 0.098 & 0.037 & 0.058 & 0.050 & 0.071 \\
J1909--3744  & 0.374 & 0.094 & 0.097 & 0.622 & 0.484 & 0.246 & 0.145 & 0.123 & 0.063 & 0.143 & 0.128 & 0.062 \\
J1933--6211  & 0.116 & -0.062 & 0.056 & 0.165 & 0.105 & 0.037 & -0.070 & -0.058 & -0.033 & 0.059 & 0.056 & 0.054 \\
J1939+2134  & 0.319 & 0.005 & 0.012 & 0.395 & 0.314 & 0.245 & 0.022 & 0.005 & -0.003 & 0.016 & 0.009 & 0.021 \\
J1944+0907  & 0.094 & 0.096 & 0.047 & 0.075 & 0.111 & 0.009 & 0.122 & 0.086 & -0.001 & 0.052 & 0.040 & 0.049 \\
J2051--0827  & 0.114 & 0.035 & 0.061 & 0.042 & 0.085 & 0.215 & 0.050 & 0.018 & 0.084 & 0.065 & 0.057 & 0.065 \\
J2124--3358  & 0.235 & -0.035 & 0.017 & 0.378 & 0.194 & 0.099 & -0.022 & -0.0356 & -0.042 & 0.011 & 0.019 & 0.058 \\
J2129--5721  & 0.516 & -0.279 & 0.248 & 0.671 & 0.459 & 0.082 & -0.263 & -0.288 & -0.052 & 0.309 & 0.224 & 0.054 \\
J2145--0750  & 0.140 & 0.053 & 0.059 & 0.173 & 0.150 & 0.097 & 0.070 & 0.065 & 0.021 & 0.086 & 0.061 & 0.046 \\
J2150--0326  & 0.164 & 0.017 & 0.034 & 0.184 & 0.095 & 0.014 & 0.028 & 0.036 & -0.004 & 0.035 & 0.050 & 0.059 \\
J2241--5236  & 0.125 & -0.031 & 0.044 & 0.170 & 0.121 & 0.070 & -0.017 & -0.035 & -0.027 & 0.028 & 0.046 & 0.047 \\
\hline
\end{tabular}
\end{table*}

\section{Summary and conclusions}\label{conclusions}

We present a new method for developing high-fidelity templates by modelling frequency evolution of pulse shapes of all four Stokes parameters. The technique also allows for studying polarisation properties, such as  phase- and frequency-resolved polarisation profiles, fractions, position angles, flux densities and spectral indices.

Our results can be summarised as follows:

\begin{enumerate}
    \item Comparing TOAs obtained with our new polarimetric templates and total intensity templates, we see that the improvement in TOA uncertainties is up to $\sim$20-30\%. Considering the fact that currently root-mean-square of the residuals is between several $\mu s$ to hundreds of ns (e.g., \citealp{PPTA_noise}) and the deviations induced by gravitational waves are of the order of tens of ns, reducing the measurement uncertainties even by $\sim$100-200~ns is extremely valuable. As a large fraction of PPTA pulsars is bright, has high levels of polarisation and profile evolution in all Stokes parameters, we expect polarimetric timing to be highly beneficial for the majority of them. 
    \item Correcting each observation for DM and RM prior to the analysis results in better profile alignment (in time and frequency) and reveals sharper profile features compared to the previous works.
    \item We show polarisation profiles at eight sub-bands, averaged into three sub-bands A (40~cm), E (20~cm) and H (10~cm) and averaged over the entire UWL band. The sub-banded profiles reveal often significant frequency evolution of the profile shapes and their complexity which are varying across Stokes parameters. These changes may also be different in various pulse components. For most pulsars profile shapes become narrower and simpler with increasing frequency, however there are several exceptions. 
    \item Median polarisation fractions across all pulsars are fairly stable in the three bands (slightly dropping down in the H band) and of the order of $L/I \sim 0.18 - 0.26$ and $|V|/I \sim 0.05$. 
    \item Evolution of the polarisation fractions in individual pulsars usually means depolarisation with increasing frequency. However, the phase- and frequency-resolved analysis shows that in some cases it is not a monotonic, but rather unimodal trend (e.g. PSR J0900--3144 and PSR J1017--7156). We note that observations with the UWL receiver are heavily affected by RFI in the middle of the band (2300-2500~MHz and 2600-2700~MHz) which may have impact on frequency-averaged data. However, we believe that the effect is physical as unimodal trends a) are seen in only several pulsars, b) do not depend on the frequency resolution of the portrait (we checked data divided into 8, 16, 32 and 416 sub-bands), c) have different directions and maxima/minima at various frequencies. 
    \item For the first time, we present spectra for PPTA pulsars with ultra-wide and contiguous frequency coverage. Moreover, we present first measurements of flux above 2~GHz for seven pulsars. Spectra of the majority of our sources are characterised with power laws, however six pulsars are best fit with a broken power law (eg. PSR J0900--3144, PSR J1545--4550). 
    \item Phase- and frequency-resolved polarisation fractions show that many pulsars exhibit smooth and symmetrical coevolution of fractions and profile shapes seen as characteristic arcs. 
    \item PA curves at different frequencies are very well aligned in phase which might indicate that radio emission of millisecond pulsars originates from an extremely narrow height range.
    \item We see a correlation between spin-down energy $\dot{E}$ and profile complexity (simpler profiles tend to have higher $\dot{E}$). 
\end{enumerate}
This paper demonstrates the importance of studying pulsars not only with ultra-widebandwidth receivers but also with methods adapted to handle all the information provided by large frequency coverage.


\section*{Acknowledgements}

We thank Simon Johnston for valuable discussions and feedback. Murriyang, CSIRO’s Parkes radio telescope, is part of the Australia Telescope National Facility (https://ror.org/05qajvd42) which is funded by the Australian Government for operation as a National Facility managed by CSIRO. We acknowledge the Wiradjuri people as the Traditional Owners of the Observatory site. Parts of this work was completed with support from the Australian Research
Council (ARC) Centre of Excellence CE230100016. MEL is supported by an Australian Research Council (ARC) Discovery Early Career Research Award DE250100508.
This work was performed on the OzSTAR national facility at Swinburne
University of Technology. The OzSTAR program receives
funding in part from the Astronomy National Collaborative
Research Infrastructure Strategy (NCRIS) allocation provided
by the Australian Government.



\bibliographystyle{mnras}
\bibliography{refs}


\newpage

\appendix

\section*{Data Availability}
Results presented here, including portraits, plots and code are available publicly as supplementary online data. 

\section{Notes on individual pulsars}\label{appendix:A}

\subsubsection{PSR J0030+0451}
Profile shapes of I and L coevolve in frequency. Disjoint profile component is highly polarised at lower frequencies, then depolarises, and has a shallower spectral index than the main pulse. Main pulse fractions are stable across frequencies, with a small decrease. Phase-resolved spectral index has a sharp peak at the location of most profile evolution with frequency. Disjoint profile component does not evolve significantly. PA does not seem to evolve with frequency. 

\subsubsection{PSR J0125--2327}
Constant negative sign of V in both phase and frequency. Phase-resolved index oscillates around band-averaged value between phase 0.4 and 0.6. There is an abrupt sharpening of the spectral index at the phase corresponding to the part of the pulse evolving the most. 
There are three PA jumps corresponding to sudden drops of L (visible in sub-banded profiles, not the band average). 
Peak of V shifts to lower phase values. V fraction increases to about bandwidth centre and then drops down. L increases to around 1000 MHz and then drops down. At low frequencies, L is broader and closer to pulse edges, however it shifts towards the centre of the phase with frequency. Shape of L evolves severely, especially in the main pulse component. 

\subsubsection{PSR J0437--4715}
Similarly to PSR J0125-2327, fraction of V grows towards the middle of the band and then decreases. Additionally, V changes sign at middle phase. Emission is detectable almost in the entire pulse phase. Both PA and phase-resolved spectral index have complex and highly variable changes.

\subsubsection{PSR J0613--0200}
A sudden change of sign in V in the lowest frequencies from negative to positive. Fraction of V grows with frequency, whereas L drops down. Polarisation profiles of both L and V shift in phase from right to left with increasing frequency. 
Severe profile evolution in each profile. Total intensity profile is sharper than in \cite{Dai15}. 

\subsubsection{PSR J0614--3329}
Fraction of L is nearly constant in the left most part of the pulse profile. Disjoint profile component is generally weakly polarised, however L fraction is high at the leading edge and is maintained across the whole band. There is a lot of profile evolution after phase 0.4 reflected also in increased alpha changes. 
There are sign changes of V in phase; L slightly increases up to around 1.4 GHz and then decreases with overall change by about 10\%. 

\subsubsection{PSR J0711--6830}
There is a constant negative sign of V across phase and frequency. Broad profile with three main profile components. Profiles of L and V seem to mirror each other. The amplitude of the small, rightmost profile component drops down significantly with frequency. Similarly to PSR J0614-3329, the disjoint profile component is weakly polarised, but has constant and high L fraction on the leading edge. Fractions of both L and V grow in the first profile component, but drop down in the second. When phase averaged, fraction of L drops down, V drops down towards the middle of the band and then rises again. 

\subsubsection{PSR J0900--3144}
Complexity of profile shape of all polarisation profile components grows with frequency. 
There is a visible profile micro-component at the leftmost phases, however it is below our detection threshold. The sign of V changes in both phase and frequency. 
Fraction of L increases towards the middle of the band and then decreases, whereas fraction of V increases. 

\subsubsection{PSR J1017--7156}
Total intensity profile shape (which is quite simple) does not evolve significantly with frequency, however both L and V do. 
Both fractions of L and V change non monotonically (are unimodal functions of frequency). Fraction of L increases towards the middle of the band and then drops down, V also changes but in opposite direction. Our profile is sharper than in \cite{Dai15} due to applied DM corrections. 

\subsubsection{PSR J1022+1001}
There is nearly constant fraction of L on the right side of the pulse profile. PA is constant across frequencies and follows a standard RVM swing. V is negative throughout most of the band, but part of its profile changes sign at high frequencies. Large sweep in phase-phase resolve spectral index corresponds to height exchange between profile components.

\subsubsection{PSR J1024--0719}
There is nearly constant fraction of L at the leading edge of the pulse. PA at corresponding phases is flat. 
Fraction of V fraction is small but stable across frequencies, changes sign in both phase and frequency. Profile shape of total intensity (and L) evolve significantly. 

\subsubsection{PSR J1045--4509}
There is a faint disjoint profile component with growing fraction of L with frequency. Generally, there is not much of profile evolution. Fractions of L and V are the similar levels and only slightly decrease wit frequency. Phase-resolved spectral index changes are smooth (possibly because of the simple profile shape) and drop/rise at the leading/trailing edges where most of the profile evolution occurs. 

\subsubsection{PSR J1125--6014}
There is faint bridge emission in the long disjoint profile component. Its two profile components evolve differently, with high fraction of L especially in the lower frequencies. Trailing edge of one of the disjoint profile components has nearly constant fraction of L. 
Fraction of V is very small. Shape of L evolves significantly. There is a PA jump which can be associated with disappearing of L at higher frequencies. We also note a different tempo of depolarisation in each profile component.  

\subsubsection{PSR J1446--4701}
This pulsar has a simple profile, hardly evolving with frequency. Fraction of L higher near the leading edge. Profile shapes of L and V seem to mirror each other. Spectral index steepens with decreasing L. Both L and V drop down. There is a simple PA sweep. 

\subsubsection{PSR J1545--4550}
We note two profile micro-components on both sides of the main pulse. Both seem to have the some fraction of L, but it is below our detection limit. The smaller profile component of the main pulse is highly polarised in L throughout half of the band. Leading edge has a bump which disappears at higher frequencies. Profile of V in the main pulse evolves from negative to suddenly positive at high frequencies. Spectral index peaks align with most frequency evolving parts of the profile. 
Fractions of L and V increase towards the centre of band and then decrease. 
PA is stable across frequencies and flat which is also related to high fraction of L.
Previously reported profiles were missing at 40cm/50cm due to low S/N \citep{Dai15}. 

\subsubsection{PSR J1600--3053}
Profile shape of L is much more complicated than I. There is a sign change in V. 
We note a large sweep in spectral index of the most evolving profile component. 
Both fractions of L and V first increase and then decrease. Quite simple non evolving PA. There seems to be a very faint micro-component emerging at phases $\sim$0.15-0.25 in the lowest frequency bin, however it is below our detection threshold.  

\subsubsection{PSR J1603--7202}
This pulsar has a very high fraction of V. There is some elongated faint emission on both sides of the main pulse. It makes a bridge to the disjoint profile component, which is highly polarised, has much shallower spectral index, flat PA, and its amplitude increases with frequency. Micro-emission on the left of the main pulse also seems to be linearly polarised and has a flat PA at corresponding phases.

\subsubsection{PSR J1643--1224}
This pulsar has a simple profile, however affected by scattering. There are changes sign of V in phase and frequency. Profiles of L and V are more complicated than I and highly evolving. 
Spectral index has the sharpest change corresponding the the scattering tail. 
Fractions of L and V are unimodal functions of frequency.

\subsubsection{PSR J1705--1903}
This pulsar has a very sharp, simple profile, affected by scattering. Phase-resolved spectral index changes significantly, which can be associated with scattering. 

\subsubsection{PSR J1713+0747}
In this pulsar, polarisation fraction of L does not evolve much with frequency in the middle of the main peak and on the trailing edge. The leading edge stays highly polarised until the middle of the band. Profile shape does not evolve significantly except from the second from the right pulse component. There is a sign change in V. 
This pulsar has been analysed recently in detail in \cite{Mandow25}.

\subsubsection{PSR J1730--2304}
This pulsar has a quite complicated profile with a similar level of complexity in each band. Fraction of L is higher towards the edges of the profile and does not change much with frequency. 
Disjoint profile component is highly polarised in L and has flat PA. The leading edge of the main component is elongated and consists of 3 micro-components, where two are highly polarised and the middle one is polarised on its trailing edge only. 

\subsubsection{PSR J1744--1134}
This pulsar is characterised by the highest fraction of L in our sample. It starts decreasing only above 3000 MHz. There are small spectral index oscillations, which might be associated with phases where L changes at high frequencies. 
Steep sweep of PA at the pulse centre. Disjoint profile component is visible across the whole band, however it has little to no polsarisation. 

\subsubsection{PSR J1824--2452A}
This pulsar has a complex profile with very high fraction of L; PA is flat. 
Left profile component has much steeper spectral index and lower fraction of L which increases with frequency. Rightmost profile component, which is connected to the central one, disappears with frequency and has high fraction of L on the trailing edge. 
Leftmost profile component has two parts, with different frequency-evolution trends. There is a detectable fraction of V only in the leftmost profile component. 

\subsubsection{PSR J1832--0836}
This pulsar has a complex multi-component profile. Profile components at the outer edges have high fraction of L maintained across frequencies. 

\subsubsection{PSR J1857+0943}
This pulsar is characterised by complex profile shapes of L and V (much more than I). Changes of spectral index can be associated with the most evolving pulse edges. Fractions are quite stable across frequencies. 

\subsubsection{PSR J1902--5105}
This pulsar is very weak above 2 GHz. Interestingly, there is almost no L, but instead prominent V of opposite signs in the two main profile components. 

\subsubsection{PSR J1909--3744}
This pulsar has a simple, but sharp PA shape which is stable with frequency. Spectral index sweeps around pulse edges and peaks with highest fraction of L. There is clear depolarisation with frequency. Profile micro-component on the left is visible at all frequencies. There is a previously missed profile component closely following the main pulse. It is highly polarised in L. 

\subsubsection{PSR J1933--6211}
This pulsars has a very complex profile which simplifies at higher frequencies. It has complex PA; V changes sign. Interestingly, spectral index changes are quite simple compared to profile evolution. Polarisation fractions are quite stable.

\subsubsection{PSR J1939+2134}
We find a new sharp feature at the trailing edge of the main pulse. Profiles reported previously (e.g. \citealp{Dai15, Spiewak22}) are smooth, however this might be due to the incorrect DM and profile smearing. 
There are two profile micro-components on each left side of the main pulse that grow amplitude with frequency. The is complex evolution of the shape of L. Fraction of L decreases in the left and increases in the right profile component. We note very sharp drop downs of spectral index for both profile components, with no oscillations. They are most likely correlated with a very high DM for this pulsar, and possibly residual misalignment of the average portrait. 

\subsubsection{PSR J1944+0907}
This pulsar has a very broad profile. Fractions of L and V are on a similar level and stay stable across frequencies. Spectral index is the most stable from the whole sample of PPTA pulsars. There is not much profile evolution.

\subsubsection{PSR J2051--0827}
This pulsar has a relatively simple profile, but evolving quite significantly (sharpening and losing complexity with frequency). There is a micro-component at the lowest frequency on the left side. There is change of sign of V. Fractions are quite stable; L is slightly increasing. 

\subsubsection{PSR J2124--3358}
This pulsar has a very broad and complex profile. Two of the small profile components stay nearly 100\% polarised. There are large oscillations of the spectral index, whereas PA is simple. 

\subsubsection{PSR J2129--5721}
In this pulsar PA evolves with frequency. A new peak in PA appears in the 20cm band (close to phase 0.5) which can be associated with appearance of a sharp V peak. 

\subsubsection{PSR J2129--5721}
This pulsar has a simple PA. There are large fractions of both L and and V. Shapes of L and V evolve significantly.

\subsubsection{PSR J2145--0750}
For this pulsar, the small left profile component is highly polarised and has a steeper spectrum than the main profile component. It also disappears with frequency. There is almost no profile evolution with frequency. Fractions of L and V are quite stable. There is a faint micro-component following the main pulse, which was first reported by \cite{CPTApol}. It is visible in our data set in the three lowest frequency bands, however it can be missed if not zoomed-in. Interestingly, waterfall plots reveal the component immediately, showing approximately similar fractions of L and V. 

\subsubsection{PSR J2241--5236}
This pulsar has a complex shape of L especially when compared with a simple sharp peak of I. Flux is flattening at higher frequencies. We note quite stable fractions and little profile evolution with frequency.


\bsp	
\label{lastpage}
\end{document}